\documentclass[pdftex,letterpaper,10pt]{article}

\pdfoutput=1
\usepackage[dvips]{graphicx}
\usepackage{mathptmx}
\usepackage{amsmath}
\usepackage{latexsym}
\usepackage{amssymb}
\usepackage{abstract}
\usepackage{enumerate}
\usepackage[dvips]{color}
\usepackage{titlesec}
\usepackage{abstract}
\usepackage[square,sort&compress]{natbib}
\usepackage{ctable}
\usepackage[top=1.5in, bottom=1.5in, left=1.5in, right=1.5in]{geometry}
\newcommand{\footnoteremember}[2]{\footnote{#2}\newcounter{#1}\setcounter{#1}{\value{footnote}}}
\newcommand{\footnoterecall}[1]{\footnotemark[\value{#1}]}

\begin{document}

\title{A statistical study of the properties of large amplitude whistler waves and their association with few eV to 30 keV electron distributions observed in the magnetosphere by Wind}

\author{L.B. Wilson III\footnoteremember{1}{Department of Physics and Astronomy, University of Minnesota, Minneapolis, MN} \footnoteremember{2}{NASA Goddard Space Flight Center, Greenbelt, Maryland, USA.}, C.A. Cattell\footnoterecall{1}, P.J. Kellogg\footnoterecall{1}, J.R. Wygant\footnoterecall{1}, K. Goetz\footnoterecall{1}, \\ A. Breneman\footnoterecall{1}, and K. Kersten\footnoterecall{1}}
\maketitle


\begin{abstract}
    We present a statistical study of the characteristics of very large amplitude whistler waves inside the terrestrial magnetosphere using waveform capture data from the Wind spacecraft as an addition of the study by \citet{kellogg10b}.  We observed 244(65) whistler waves using electric(magnetic) field data from the Wind spacecraft finding $\sim$40$\%$($\sim$62$\%$) of the waves have peak-to-peak amplitudes of $\geq$50 mV/m($\geq$0.5 nT).   We present an example waveform capture of the largest magnetic field amplitude ($\gtrsim$8 nT peak-to-peak) whistler wave ever reported in the radiation belts.  The estimated Poynting flux magnitude associated with this wave is $\gtrsim$300 $\mu$W/m$^{2}$, roughly four orders of magnitude above previous estimates.  Such large Poynting flux values are consistent with rapid energization of electrons.  The majority of the largest amplitude whistlers occur during magnetically active periods (AE $>$ 200 nT).  The waves were observed to exhibit a broad range of propagation angles with respect to the magnetic field, 0$^{\circ}$ $\leq$ $\theta{\scriptstyle_{kB}}$ $<$ 90$^{\circ}$, which showed no consistent variation with magnetic latitude.  These results are inconsistent with the idea that the whistlers are all generated at the equator, propagating along the magnetic field, and that the observed obliqueness is due to propagation effects.  We also identified three types of electron distributions observed simultaneously with the whistler waves including beam-like, beam/flattop, and anisotropic distributions.  The whistlers exhibited different characteristics depending on the observed electron distributions.  For instance, the whistlers observed with anisotropic distributions in the radiation belts had larger $\Delta$f/f than the rest of the whistlers.  The majority of the waveforms observed in our study have f/f${\scriptstyle_{ce}}$ $\leq$ 0.5 and are observed primarily in the radiation belts simultaneously with anisotropic electron distributions.
\end{abstract}


\section{Introduction}  \label{sec:introduction}
\indent  Whistler waves are one of the most ubiquitous wave modes in plasmas.  They have been observed in the magnetosphere \citep{burtis69a, russell69a, parrot03a, santolik03a, cattell08, cully08a}, in the solar wind \citep{neubauer77a, breneman10a}, upstream of interplanetary shocks \citep{wilsoniii09a}, upstream of planetary bow shocks \citep{hoppe81a, bertucci05a}, and in cometary foreshocks \citep{tsurutani87}.  Whistler waves are a right-hand polarized electromagnetic mode that can propagate along the magnetic field or at highly oblique angles as a quasi-electrostatic mode near the resonance cone interacting with both ions and electrons.  Whistlers can be driven unstable by electron temperature anisotropies \citep{kennel66, hashimoto81a}, electron heat flux \citep{gary94}, and electron beams \citep{tokar84, zhang93a, sauer10a}.  For simplicity, we include chorus, plasmaspheric hiss, and oblique whistler-modes in our use of the term whistler wave.  Because whistlers interact strongly with energetic particles \citep{kennel66, lyons72a}, it has been well accepted that they play an important role in global radiation belt dynamics.  Thus, whistler waves have been a topic of extreme interest for over 40 years in magnetospheric physics.  \\
\indent   Time-averaged spectral intensities have been used to estimate whistler wave amplitudes for over 30 years \citep{gurnett76a}, but averaging can underestimate the instantaneous wave amplitudes.  Typical whistler time-averaged amplitudes are $\sim$0.5 mV/m for the electric field \citep{meredith01a} and $\sim$0.01-0.1 nT for the magnetic field \citep{horne03a, horne05a}.  The underestimated wave amplitudes may mask important wave-particle interactions which require larger amplitude waves.  Recent observations by \citet{santolik03a}, \citet{cattell08}, \citet{cully08a}, and \citet{lecontel09a} showed that previous time averaged spectral intensity measurements may often underestimate the wave amplitude of terrestrial whistler waves by more than two orders of magnitude.  Test particle simulations have found that electrons can be accelerated to MeV energies in relatively short times by these very large amplitude whistler waves \citep{omura07a, cattell08, bortnik08a}.  These observations have raised new questions regarding the energization and lifetime of radiation belt particles.  \\
\indent  During an eight year period, the Wind spacecraft went through a number of petal orbits into the terrestrial magnetosphere.  We report on whistler wave statistics for 13 of those petal orbits, building on the study by \citet{kellogg10b}, who used an automated search algorithm to find large amplitude whistlers.  A search by eye yielded significantly more whistler waves than reported in \citet{kellogg10b}.  This is the first study to examine the statistics of the relationship between wave propagation angle and magnetic latitude for these very large amplitude whistlers, finding no correlation between propagation angle and magnetic latitude.  This is the first observation of very intense whistler Poynting flux.  We also present the first statistical study of the characteristics of the electron distributions over the energy range of a few eV to 30 keV associated with large amplitude whistler waves inside the terrestrial magnetosphere using waveform capture data simultaneously with three different types of electron distributions.  The paper is organized as follows:  Section \ref{sec:data} introduces and outlines the data sets and analysis techniques, Section \ref{sec:observations} describes the observations, and Section \ref{sec:theory} discusses the results and conclusions of our study.
\section{Data Sets and Analysis}  \label{sec:data}
\indent  Waveform captures were obtained from the Wind/WAVES instrument \citep{bougeret95}, using the time domain sampler (TDS) receiver, which provides a waveform capture (herein called TDS sample) of 2048 points with timespans ranging from $\sim$17 ms to $\sim$1000 ms, depending on sample rate.  TDS samples are utilized from the fast (TDSF) and slow (TDSS) TDS receivers.  The TDS receivers have a low frequency cutoff of roughly 3.3 Hz for TDSS and 130 Hz for TDSF.  The TDS buffer stores and evaluates waveforms based upon their amplitude, keeping only the largest events.  TDSF samples are composed of two electric field vectors in the XY-GSE plane, defined as E${\scriptstyle_{j}}$.  TDSS samples are composed of four vectors, either three electric (E${\scriptstyle_{j}}$) and one magnetic (B${\scriptstyle_{j}}$) or three magnetic and one electric field components.  The TDSS samples are rotated into magnetic field-aligned coordinates (FACs), determined with magnetic field measurements from the Wind magnetic field instrument (MFI) \citep{lepping95}, where the subscript $\parallel$($\perp$) refer to parallel(perpendicular) to the magnetic field.  We define the wave amplitude as $\mid$\textbf{E}${\scriptstyle_{w}}$$\mid$ for electric fields and $\mid$\textbf{B}${\scriptstyle_{w}}$$\mid$ for magnetic fields.  The Wind/WAVES instrument also contains an onboard time-averaged spectral intensity instrument, the thermal noise receiver (TNR), used to analyze low amplitude signals and determine the local plasma density using the plasma line when possible.  Note that we only use the TNR receiver for determination of the local plasma line, not for whistler wave identification.  \\
\indent  The wave vector, \textbf{k}, and propagation angle with respect to the magnetic field, $\theta{\scriptstyle_{kB}}$, were determined using Minimum Variance Analysis (MVA) \citep{khrabrov98} on bandpass filtered TDSS samples with three magnetic field components.  The frequency ranges for each bandpass filter, determined from spectral analysis, were chosen independently for each TDS sample.  We required intermediate to minimum eigenvalues of the spectral matrix, $\lambda{\scriptstyle_{int}}$/$\lambda{\scriptstyle_{min}}$, to satisfy the condition $\geq$ 10.0 if less than 50 field vectors were used in the analysis \citep{khrabrov98}.  \\
\indent  Full 4$\pi$ steradian particle distributions for electrons and ions were obtained from the Wind/3DP EESA and PESA particle detectors for low energies ($<$30 keV) and the solid state telescopes (SSTs) for high energy ($\sim$30-500 keV) electrons (SST Foil) and ($\sim$70-6000 keV) protons (SST Open) \citep{lin95}.  Electron distribution functions, from a few eV to 30 keV, were examined for pitch-angle anisotropies as a possible free energy source.  Higher energy particle distributions using the 3DP solid state telescope (SST) detectors were used to identify the periods when Wind entered the radiation belts.  However, since the detectors were built for solar wind use, they often saturate in the radiation belts.  Thus, they could not be used to look for potential free energy sources or for correlations between flux enhancements/losses and strong waves.  We use the MFI data, combined with PESA data, to define the region we call the outer magnetosphere as the region between the terrestrial magnetopause and the radiation belts.  The magnetopause was defined by the sharp gradient in flow speed, density, B${\scriptstyle_{z}}$-GSM, and magnetic field turbulence.  We defined the radiation belt region by the sharp increase(decrease) in omni-directional flux of the high energy electrons ($>$ 100 keV) and protons ($>$ 1 MeV) as Wind approached(departed from) perigee in its petal orbits.  To allow for the possibility of different sources of free energy and/or damping, outer magnetospheric and radiation belt whistlers were examined separately.  \\
\indent  We also examined electron temperature anisotropies (T${\scriptstyle_{\perp j}}$/T${\scriptstyle_{\parallel j}}$ $\neq$ 1) for both the Eesa Low (j $=$ c) and Eesa High (j $=$ h) separately to compare to the threshold criteria for whistler anisotropy instability \citep{kennel66}.  The two EESA detectors have energy ranges ($\lesssim$30 keV) well below the typical $>$100 keV radiation belt particles.  Thus, using these detectors limits our ability to accurately describe the full particle distributions in the radiation belts.  However, we will show that many of the observed whistlers are in resonance with electrons below 30 keV.  Recall that we mentioned that the SST detectors often saturate in the radiation belts, thus we are limited to $\lesssim$30 keV electron distributions in this study.  \\
\section{Observations}  \label{sec:observations}
\begin{figure}[htb]
 \begin{center}
   {\includegraphics[trim = 0mm 0mm 0mm 0mm, clip, width=13cm]{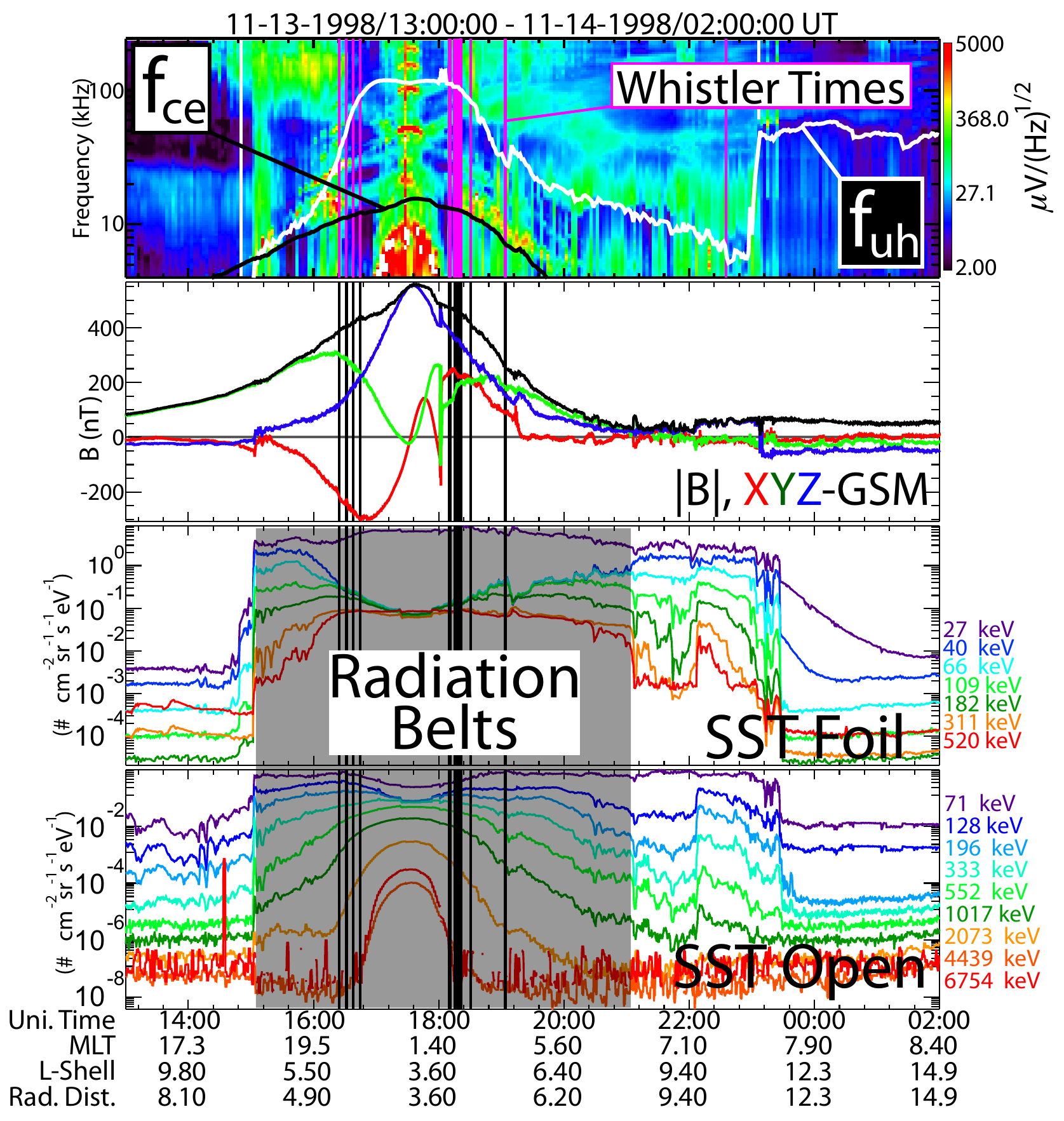}}
   \caption[]{Example of a Wind petal orbit for the time range of 1998-11-13/13:00:00 UT through 1998-11-14/02:00:00 UT.  The top panel is a time averaged spectral intensity plot from the WAVES/TNR receiver with the local f${\scriptstyle_{ce}}$ (black line) and f${\scriptstyle_{uh}}$ (white line) over plotted.  The second panel is the GSM components of the magnetic field and magnitude.  The third and fourth panels are omni-directional number fluxes of high energy electrons and protons, respectively, from the Wind SST instruments.  At the bottom of the plot are tick mark labels of MLT, L-Shell, and radial distance along with the local UT.}
   \label{fig:exWindpetalorbit}
 \end{center}
\end{figure}

\begin{figure}[htb]
 \begin{center}
   {\includegraphics[trim = 0mm 0mm 0mm 0mm, clip, width=13cm]{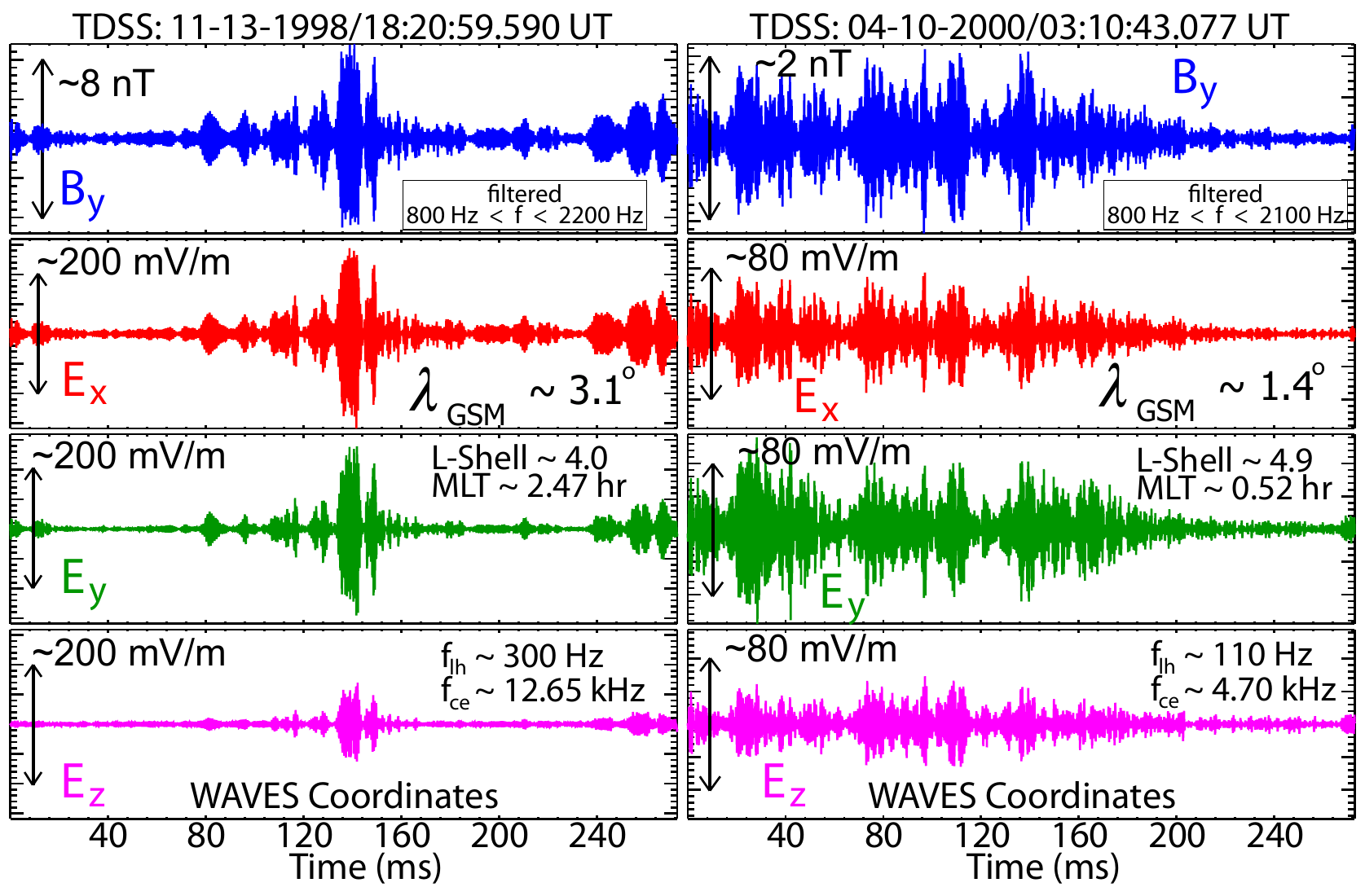}}
   \caption[]{Examples of two whistler waves, in instrument coordinates, observed on the perigee passes of Wind on 1998-11-13 and 2000-04-10.  The examples were taken from the TDSS instrument and filtered (frequency range shown in top panels) to remove superposed low and high frequency signals.  The top row shows the Y-component of the magnetic field (blue), second row shows the X-component of the electric field (red), third row shows the Y-component of the electric field (green), and the fourth row shows the Z-component of the electric field (magenta).  The amplitudes of each component are marked by the vertical black arrows.  The TDSS samples for these two events were taken at 7.5 kHz.}
   \label{fig:whistlers}
 \end{center}
\end{figure}

\begin{figure}[htb]
 \begin{center}
   {\includegraphics[trim = 0mm 0mm 0mm 0mm, clip, width=13cm]{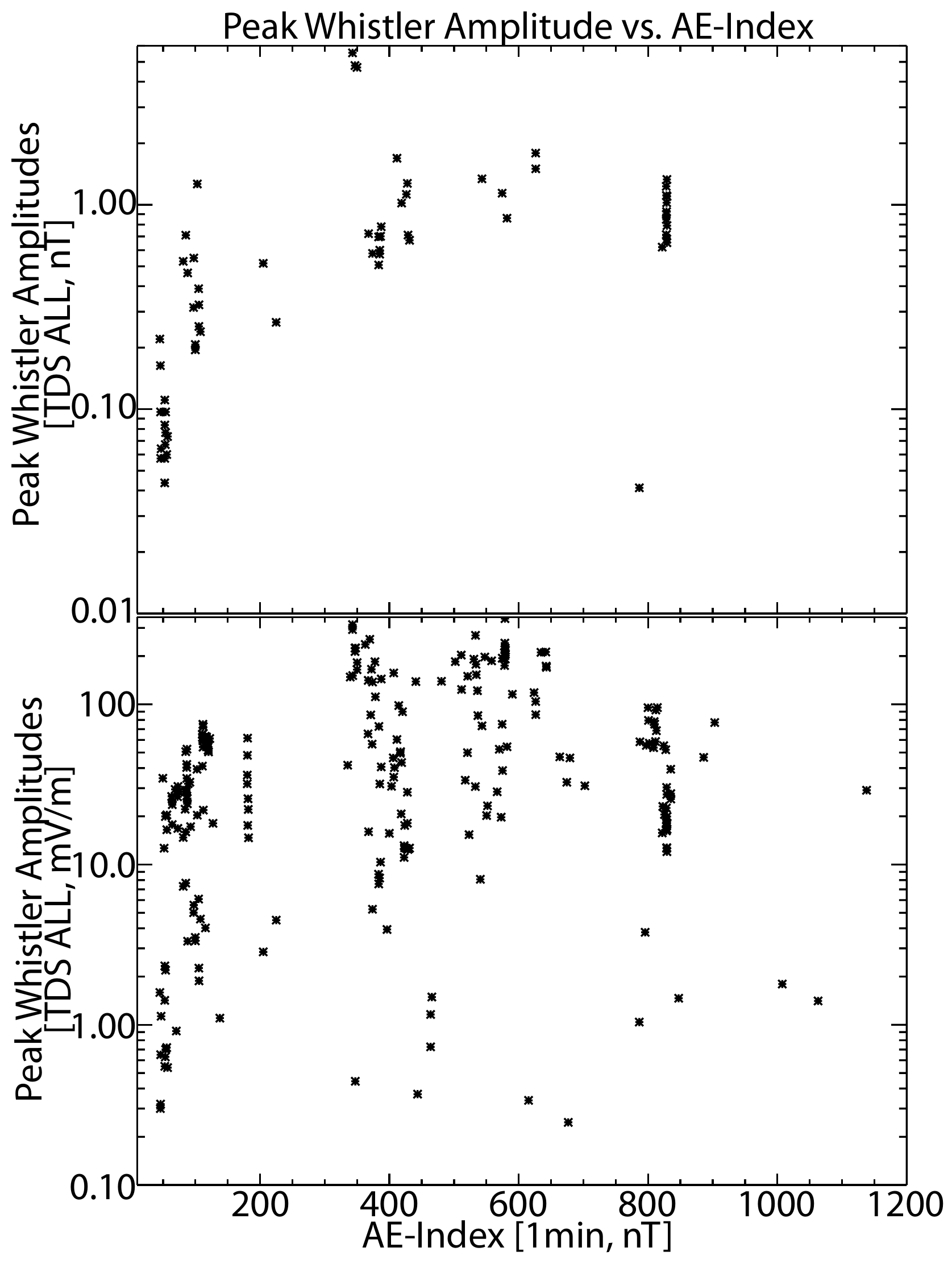}}
   \caption[]{Peak whistler wave amplitudes versus one minute AE-Indices for the whistlers.  The top panel corresponds to the peak whistler amplitudes measured with the Wind search coil (ranging from $\sim$0.01-2.0 nT) and the bottom panel is the peak whistler amplitudes measured with the Wind electric field antennas (ranging from $\sim$0.1-350 mV/m).  The range of both horizontal axes is 10-1200 nT for the AE-Indices.}
   \label{fig:peakampvsaeindex}
 \end{center}
\end{figure}

\begin{figure}[htb]
 \begin{center}
   {\includegraphics[trim = 0mm 0mm 0mm 0mm, clip, width=13cm]{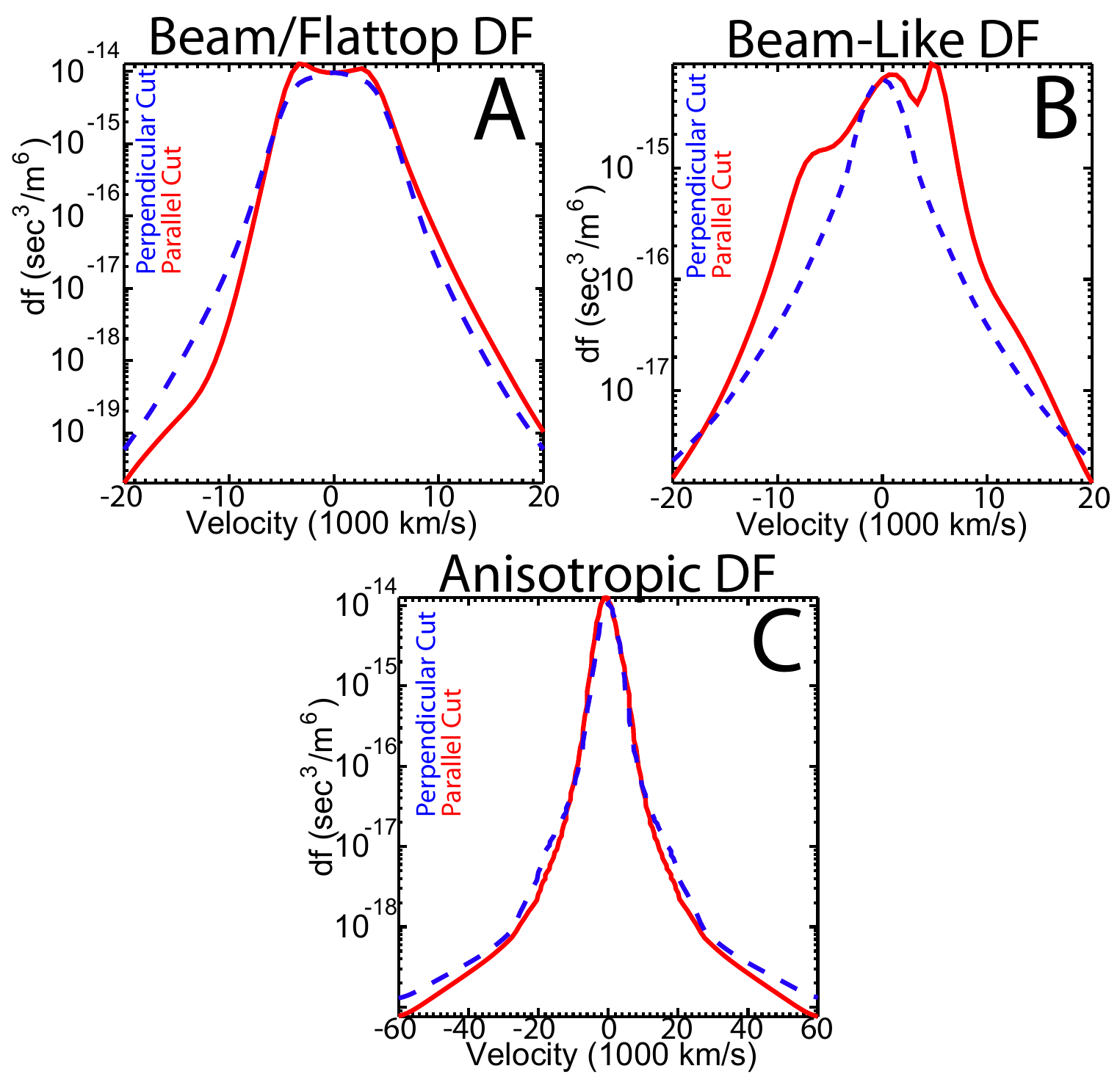}}
   \caption[]{Example cuts of electron distribution functions for beam/flattop (Type \textbf{A}), beam (Type \textbf{B}), and anisotropic (Type \textbf{C}).  The parallel(perpendicular) cuts of the electron distribution functions are seen as the red(blue) solid(dashed) lines.  The green line corresponds to the one-count level for that particular distribution.}
   \label{fig:dfELdefinitions}
 \end{center}
\end{figure}

\begin{figure}[htb]
 \begin{center}
   {\includegraphics[trim = 0mm 0mm 0mm 0mm, clip, width=10.5cm]{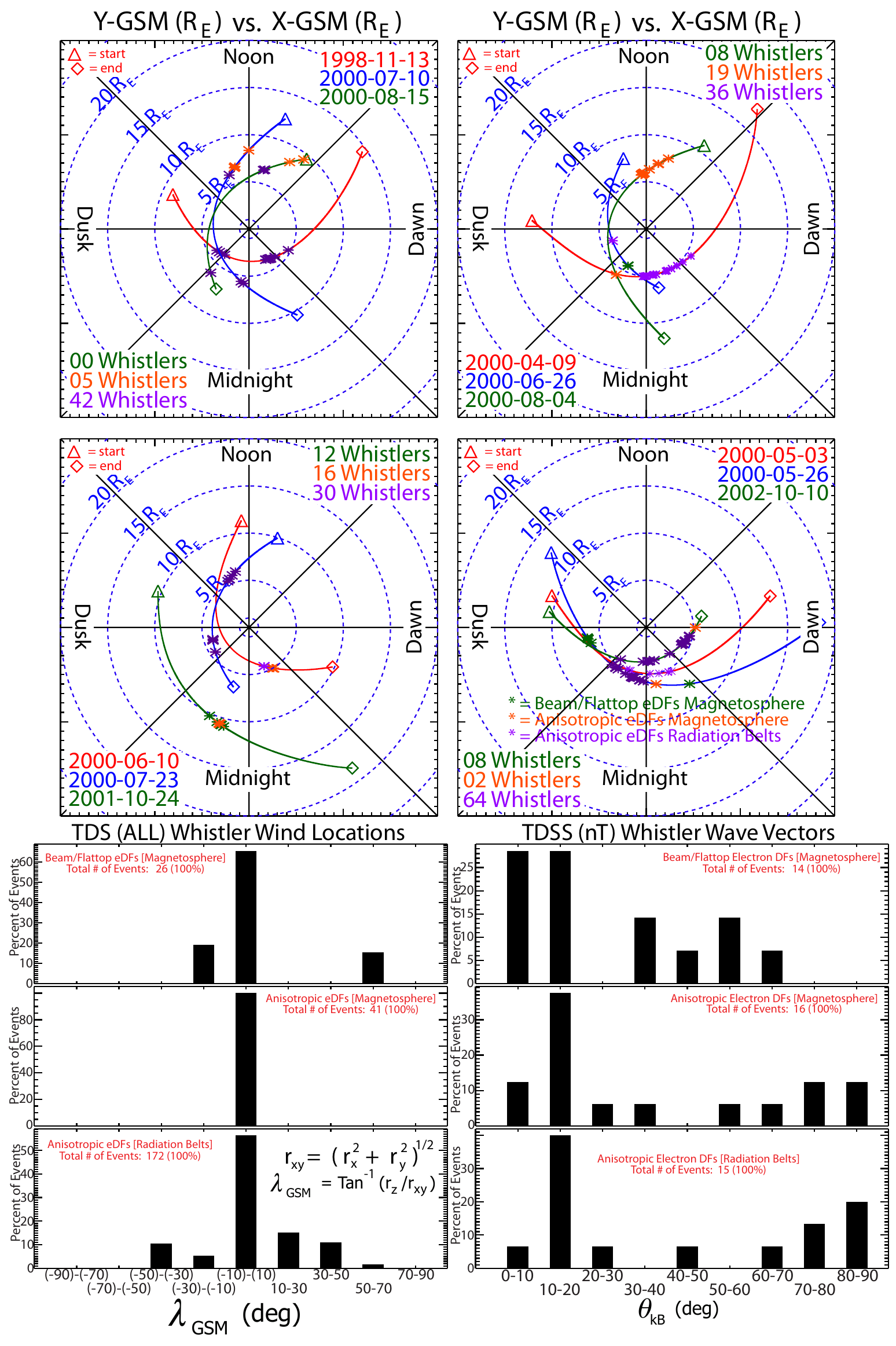}}
   \caption[]{The top four plots show 12 Wind petal orbits over layed and color coded (labeled in each panel) with whistler wave locations (defined by color coded $\ast$) projected onto the XY-GSM plane.  For each petal orbit, the start(end) point is identified by a $\triangle$($\diamondsuit$).  Below the four petal orbit plots are two histograms of GSM-Latitude, $\lambda{\scriptstyle_{GSM}}$ (degrees), for the 239 TDSF and TDSS samples on the left while on the right is the histogram of $\theta{\scriptstyle_{kB}}$ (degrees) for the 45 TDSS used in MVA.  Each histogram panel contains three histograms separating the whistler waves by the concurrent electron distribution function observations and magnetospheric location.  The plots are organized in the following order:  (top panel) beam/flattop electron distribution functions in the outer magnetosphere, (middle panel) anisotropic electron distribution functions in the outer magnetosphere, and (bottom panel) anisotropic electron distribution functions in the radiation belts.  The vertical axis is the percentage of whistlers in each event category.}
   \label{fig:PetalOrbitsLatGSMThetakB}
 \end{center}
\end{figure}

\indent  Figure \ref{fig:exWindpetalorbit} shows an example of a Wind perigee pass from 1998-11-13/13:00:00 UT to 1998-11-14/02:00:00 UT.  The top panel shows one minute averaged spectral data from the TNR receiver in $\mu$V/$\sqrt{Hz}$.  Overplotted is the upper hybrid frequency (f${\scriptstyle_{uh}}$) line.  Due to the large amounts of noise in the TNR data, the density for the f${\scriptstyle_{uh}}$ line was estimated from PESA High density measurements.  However, the PESA moments give an underestimate of density at low L-shells, so our density estimates are likely too low.  It can be seen that the whistlers generally occur between L $\sim$3-5 (locations marked by vertical lines in all panels), in the radiation belts.  Also note that the lowest energy bins of the SST detectors saturate between roughly 16:00 and 20:00 UT on 1998-11-13.  This is the most obvious in the SST Foil panel for the 40 keV line which drops in flux just after 16:00 UT.  This saturation limited our ability to use the SST detectors when examining the whistler waves.  \\
\indent  Figure \ref{fig:whistlers} plots two whistler waves observed near the equatorial plane, post midnight near L $\sim$ 4.  These whistlers were obtained from the TDSS instrument (with three electric and one magnetic field sampled at 7.5 kHz); the waveforms shown were filtered (frequency range shown in top panels) to remove superposed low and high frequency signals.  The L-shell, MLT, GSM latitude, lower hybrid frequency (f${\scriptstyle_{lh}}$ $=$ $\sqrt{f{\scriptstyle_{ce}} f{\scriptstyle_{ci}}}$), and f${\scriptstyle_{ce}}$ for each corresponding whistler are labeled in the figure.  Note that the 8 nT peak-to-peak amplitude we report here is only one component of the magnetic field.  The total magnetic amplitude for both waveforms shown in Figure \ref{fig:whistlers} is larger.  The polarization of the electric fields is elliptical for both examples.  The 1998-11-13 example is much more oblique ($\theta{\scriptstyle_{kB}}$ $\sim$ 73$^{\circ}$) and elliptical ($\lambda{\scriptstyle_{max}}$/$\lambda{\scriptstyle_{min}}$ $\sim$ 10.7) than the 2000-04-10 example ($\theta{\scriptstyle_{kB}}$ $\sim$ 39$^{\circ}$ and $\lambda{\scriptstyle_{max}}$/$\lambda{\scriptstyle_{min}}$ $\sim$ 1.9).  \citet{omura07a} gave a relationship for the maximum change in kinetic energy given by:
\begin{equation}
  \label{eq:dKenergy}
  (\Delta KE){\scriptstyle_{max}} \approx \frac{5.6 \times 10^{4}}{L^{2} \sqrt{1 + \xi{\scriptstyle_{o}}^{2} }} \frac{\delta B}{B{\scriptstyle_{o}}} \left[MeV\right]
\end{equation}
where L is the L-shell, $\xi{\scriptstyle_{o}}^{2}$ $=$ $\omega$ ($\Omega{\scriptstyle_{EQ}}$ - $\omega$)/$\omega{\scriptstyle_{pe}}^{2}$, $\Omega{\scriptstyle_{EQ}}$ is the equitorial cyclotron frequency, and $\delta$B/B${\scriptstyle_{o}}$ is the ratio of wave amplitude to background magnetic field.  For the waves shown in Figure \ref{fig:whistlers}, ($\Delta$KE)${\scriptstyle_{max}}$ $\sim$ 61 MeV(28 MeV) for the wave on 1998-11-13(2000-04-10).  The examples show that:  (1)  large amplitude whistler waves in the radiation belts are bursty; (2) electric fields are in excess of two orders of magnitude above previous estimates; (3)  the first observations of these very large amplitude whistler waves with search coil magnetic fields shows amplitudes exceeding two orders of magnitude above previous estimates; and (4) the waves are capable of producing electrons with energies greater a MeV.  \\
\indent  To illustrate the occurrence rate of these large amplitude whistlers we will discuss two Wind perigee passes in detail.  During the perigee passes of 1998-11-13 and 2000-04-10, Wind observed some of the largest whistler events ever recorded in the magnetosphere.  On 1998-11-13 between 18:15:08.170 UT and 18:21:17.884 UT, or roughly 6 minutes, Wind observed 11 TDSF and 3 TDSS samples.  All 14 TDS samples had $\mid$\textbf{E}${\scriptstyle_{w}} \mid$ $\geq$ 100 mV/m and all 3 TDSS samples had $\mid$\textbf{B}${\scriptstyle_{w}} \mid$ $\geq$ 4 nT.  The large number of whistlers observed during such a short duration and spatial region is consistent with STEREO observations \citep{cattell08}.  On 2000-04-10 between 02:45:39.736 UT and 04:32:44.807 UT, or roughly 1.75 hours, the spacecraft moved through $\sim$3.48 hr($\sim$52.2$^{\circ}$) of MLT and $\sim$0.55 L-shell.  During this interval, Wind observed 34 TDS samples, 27 TDSF and 7 TDSS.  Of those 34 samples, 31 had $\mid$\textbf{E}${\scriptstyle_{w}} \mid$ $\geq$ 80 mV/m and all 7 TDSS samples had $\mid$\textbf{B}${\scriptstyle_{w}} \mid$ $\geq$ 1 nT.  \\
\indent  Since we only have four field components for each TDSS sample, we cannot fully describe the wave Poynting flux.  However, we can calculate part of two components to estimate the magnitude of the Poynting flux for these waves.  The 1998-11-13(2000-04-10) whistler in Figure \ref{fig:whistlers} has Poynting flux magnitude of $\gtrsim$300 $\mu$W/m$^{2}$($\gtrsim$30 $\mu$W/m$^{2}$), which is roughly four(three) orders of magnitude larger than the estimates found by \citet{santolik10a}.  To estimate the possible impact of the large Poynting fluxes, we perform the calculation of \citet{santolik10a} assuming the same estimates of $\gtrsim$1 MeV electron fluxes, background densities, and field-aligned column area (of a flux tube).  We find that it would take roughly 5 ms(50 ms) for the whistlers seen in Figure \ref{fig:whistlers} to deposit the necessary energy density ($\sim$10$^{-4}$ J/m$^{2}$) to accelerate plasma sheet electrons to 1 MeV, assuming 100$\%$ efficiency, in the outer radiation belt.  If we now assume a 1$\%$ efficiency, we find a time scale of 0.5 seconds(5.5 seconds) necessary for these whistlers to produce the same effect.  These estimates are five to six orders of magnitude shorter than the typical estimates.  However, as one can see in Figure \ref{fig:whistlers}, these whistlers are very bursty and their amplitudes are not sustained for longer than 10's of milliseconds.  We should also note that the larger amplitude waves we observe will likely not interact with electrons in the same quasi-linear fashion as described by \citet{santolik10a}.  \\
\indent  Figure \ref{fig:peakampvsaeindex} shows the relationship between peak whistler wave amplitudes and one minute AE-Index for the whistlers.  The plots show a slight trend of increasing peak amplitudes with increasing AE, consistent with increased substorm activity providing more free energy for whistler growth.  A cursory examination of LANL geosynchronous low energy (30-300 keV) electron data shows large injections on the nightside just prior to the two whistlers observed in Figure \ref{fig:whistlers}, consistent with the AE-index analysis.  \\
\indent  Figure \ref{fig:dfELdefinitions} illustrates the different types of electron distribution functions observed at times close to and/or concurrently with the whistlers.  Each plot represents the parallel and perpendicular cuts of the electron distribution function seen as the solid red and dashed blue lines, respectively.  The first panel, Type \textbf{A}, shows beam-like and/or flattop features in the parallel cut, called a beam/flattop electron distribution function.  The second panel, Type \textbf{B}, does not show a flattop feature, but does show beam-like features.  Finally, the third panel, Type \textbf{C}, is an example of an electron distribution function that has T${\scriptstyle_{\perp}}$/T${\scriptstyle_{\parallel}}$ $>$ 1.0 for $\sim$400 eV to $\geq$30 keV, called an anisotropic electron distribution function.  The anisotropy seen in Type \textbf{C} extended well beyond 1 keV, while the characteristic features of Types \textbf{A} and \textbf{B} occurred primarily below 1 keV.  We will refer to whistlers seen near Types \textbf{A} and \textbf{B} as beam/flattop-related whistlers and whistlers seen near Type \textbf{C} as anisotropy-related whistlers.  Note that due to the differences in sampling times ($\sim$3 seconds for electron distribution functions versus $\sim$17-250 ms for whistlers) there may be short-lived features in the electron distribution functions that are not observed.  Table \ref{tab:whistlerstats} shows that most of the whistlers were observed in the radiation belts and were associated with anisotropic electron distribution functions.  \\
\indent  Figure \ref{fig:PetalOrbitsLatGSMThetakB} summarizes the whistler wave location and wave vector direction statistics.  The figure shows the twelve Wind orbits in the top four panels projected onto the XY-GSM plane with color-coded lines and the date for each line is labeled in the corresponding panel (\textit{e.g.} red line for 1998-11-13 petal orbit in top left panel).  The plots of the petal orbits illustrate the limited coverage of the magnetosphere by Wind.  Over plotted on each orbit are color-coded symbols, $\ast$'s, where the colors green, orange, and purple respectively correspond to the following electron distribution functions:  (1) outer magnetospheric beam/flattop, (2) outer magnetospheric anisotropic, and (3) radiation belt anisotropic electron distribution functions.  For reference, the cumulative number of each type of whistler seen in each panel are labeled for corresponding color-coded $\ast$'s (\textit{e.g.} 5 whistlers seen with anisotropic electron distribution functions in the outer magnetosphere in top left panel).  \\
\indent  The bottom part of Figure \ref{fig:PetalOrbitsLatGSMThetakB} shows two sets of histograms of GSM-Latitude of Wind, $\lambda{\scriptstyle_{GSM}}$, on the left and $\theta{\scriptstyle_{kB}}$, the whistler wave vector angle, on the right.  The histogram for $\lambda{\scriptstyle_{GSM}}$ includes 239 TDSF and TDSS samples and shows the corresponding range of latitude during the four petal orbits.  Since $\theta{\scriptstyle_{kB}}$ could only be determined for TDSS events with three magnetic field components and we required that $\lambda{\scriptstyle_{int}}$/$\lambda{\scriptstyle_{min}}$ $\geq$ 10.0 be satisfied, we determined $\theta{\scriptstyle_{kB}}$ for only 45 whistlers.  Roughly 95$\%$(65$\%$) of the whistlers were observed within 2 R${\scriptstyle_{E}}$(10$^{\circ}$) of the XY-GSM plane largely due to the Wind spacecraft's orbits.  All of the outer magnetospheric whistlers for both electron distribution function types are observed between 4.5 $\leq$ L $\leq$ 12 and all radiation belt whistlers are observed between 3 $\leq$ L $\leq$ 7.5.  Also, 92$\%$ of the radiation belt whistlers and all the beam/flattop-related whistlers are observed within $\pm$6 hours of midnight, while only 44$\%$ of the anisotropic outer magnetospheric whistlers occurred in this range.  Most whistlers observed in the outer magnetosphere had wave vectors close to parallel ($\sim$60$\%$ had $\theta{\scriptstyle_{kB}}$ $\leq$ 30$^{\circ}$), while the whistlers in the radiation belts had a wider range of wave vectors (0$^{\circ}$ $\leq$ $\theta{\scriptstyle_{kB}}$ $<$ 90$^{\circ}$).  \\
\indent  Table \ref{tab:whistlerstats} shows the statistics of all the identified whistler wave samples inside L $=$ 15.  Two hundred fifty-seven whistlers were observed inside the magnetosphere; of these, 244 whistlers occurred between 3 $<$ L $<$ 12.  The top part of the table separates the whistlers into three columns by region and into two rows defined by the type of electron distribution function.  The middle part of Table \ref{tab:whistlerstats} shows the statistics for the whistler wave amplitudes.  The values in each column represent the fractional number of events with values satisfying the criteria for each column with at least one electric (top row) or magnetic (bottom row) field component.  There were 244 whistlers observed with at least one electric field component and 65 with at least one magnetic field component.  Below these ratios are the L-shell ranges over which the whistlers with the corresponding amplitudes were observed.   Note that a significant fraction, $\sim$40$\%$($\sim$62$\%$), of the electric(magnetic) field whistlers have $\geq$50 mV/m($\geq$0.5 nT) amplitudes.  \\
\indent  The bottom part of Table \ref{tab:whistlerstats} shows the wave amplitudes versus AE index statistics as the mean $\pm$ the standard deviation of the mean.  The electric(magnetic) field whistlers have large wave amplitudes ($\gtrsim$110 mV/m or $\gtrsim$1.6 nT) for moderately strong AE (200-400 nT), suggesting substorm injections may provide some of the free energy for the observed whistlers.  The mean amplitudes appear to decrease in the bottom part of Table \ref{tab:whistlerstats} due to the limited number of samples for AE $\gtrsim$ 600 nT, as seen by the trend in Figure \ref{fig:peakampvsaeindex}.  Note that for 10 $\leq$ AE $\leq$ 200 nT (see Figure \ref{fig:peakampvsaeindex}), the average wave amplitudes are $\sim$0.40 $\pm$ 0.04 nT($\sim$31.54 $\pm$ 1.39 mV/m), much smaller than for AE $>$ 200 nT.  \\
\indent  Although, much of the literature focuses on the relationship between whistlers and higher energy electrons \citep{li10a}, comparison to electron distributions with energies $\leq$ 30 keV indicate that the properties of the waves vary with the shape of the distributions in this energy range.  Also, previous work has shown that a Cerenkov resonance can excite whistler waves \citep{singh72a, kumagai80a, starodubtsev99c, kellogg10b}, which has much lower energies than the typical cyclotron resonance invoked.  Note that the effect of Landau damping by electrons in this energy range on propagation angles of the whistlers has also been studied \citep{bortnik06a}.  Below we will discuss estimates of electron/whistler parallel resonance energies for a range of electron pitch angles.  The total electron energy is at a minimum for electrons with zero pitch-angle.  For instance, if a particle has a pitch-angle of 45$^{\circ}$, the total particle energy is twice the parallel energy.  We will discuss those consequences below as well, but first we will discuss just parallel resonance energies.  Although there are some uncertainties in our density estimates, as explained above, we found that most of the whistlers were observed under conditions where they were resonant with electrons within the energy range of the EESA detectors.  The well known nonrelativistic parallel resonance energy \citep{kennel66b} of a particle is given by:
\begin{equation}
  \label{eq:paracycenergy}
  E{\scriptstyle_{\parallel res}} = \Bigl( \frac{ B{\scriptstyle_{o}}^{2} }{ 2 \mu{\scriptstyle_{o}} n{\scriptstyle_{e}} } \Bigr) \Bigl( \frac{ \Omega{\scriptstyle_{ce}} }{\omega \cos^{2} {\theta{\scriptstyle_{kB}}} } \Bigr) \Bigl[ \cos{ \theta{\scriptstyle_{kB}} } - \frac{\omega}{\Omega{\scriptstyle_{ce}}} \Bigr] \Bigl[ m + \frac{\omega}{\Omega{\scriptstyle_{ce}}} \Bigr]^{2}
\end{equation}
where B${\scriptstyle_{o}}$ is the magnitude of the magnetic field, n${\scriptstyle_{e}}$ is the ambient plasma density, $\Omega{\scriptstyle_{ce}}$ is the electron cyclotron frequency, $\theta{\scriptstyle_{kB}}$ is the wave propagation angle with respect to the magnetic field, and \textit{m} $=$ 0 (Landau), $-$1 (normal cyclotron), or $+$1 (anomalous cyclotron) for the different resonances. \\
\indent  For the two examples shown in Figure \ref{fig:whistlers} the necessary parameters needed to calculate resonant energies are the following:  f $\sim$ 900-2200 Hz(800-2100 Hz), $\theta{\scriptstyle_{kB}}$ $\sim$ 73$^{\circ}$(39$^{\circ}$), N${\scriptstyle_{i}}$ $\sim$ 168 cm$^{-3}$(71 cm$^{-3}$), and $\mid$\textbf{B}$\mid$ $\sim$ 452 nT(168 nT) for the waveform on the left(right).  Given the above parameters, the Landau resonant energies from the cold plasma dispersion relation are $\sim$ 560-720 eV(170-240 eV) and the normal cyclotron resonant energies are $\sim$ 16.2-96.2 keV(0.37-4.02 keV) for the waveform on the left(right).  Of course if our estimates of either the density or propagation angle are wrong we receive slightly different results.  We know that the PESA High detector saturates in the radiation belts and under estimates the densities.  Thus, increasing the density lowers both the Landau and normal cyclotron resonance energies for both the whistler at 1998-11-13/18:20:59.590 UT and the one at 2000-04-10/03:10:43.077 UT.  Lowering the propagation angle has the same effect, in the cold plasma limit with all other parameters in Equation \ref{eq:paracycenergy} being constant, since E${\scriptstyle_{\parallel res}}$ is roughly proportional to 1/$\cos{\theta{\scriptstyle_{kB}}}$.  \\
\indent  Particles with a nonzero pitch-angle have a higher total energy than given by Equation \ref{eq:paracycenergy}.  This is important because the detectors measure the total energy of a particle in a specific angular bin, not just the parallel energy.  Thus, if the particles have large pitch-angles and E${\scriptstyle_{\parallel res}}$ $\lesssim$ 30 keV, it is possible that the detector would not measure the particle.  This becomes an issue at high pitch-angles ($\gtrsim$45$^{\circ}$) which encompasses a significant fraction of the total distribution when anisotropic.  However, the waves we observe are highly oblique (as seen below) and thus have large longitudinal electric fields.  The consequence is that the Landau resonance energies may be more important than the cyclotron resonance energies \citep{kennel66, kumagai80a, starodubtsev99c, kellogg10b}.  Thus, we argue that the distributions produced by the EESA detectors can be relevant to the observed whistler waves.  \\
\indent  Recent simulation work by \citet{sauer10a} found that oblique whistler waves can be excited by electron beams if the beam speed, V${\scriptstyle_{b}}$, is at least greater than twice the electron Alfv$\acute{e}$n speed, V${\scriptstyle_{Ae}}$ $=$ B${\scriptstyle_{o}}$/$\sqrt{\mu{\scriptstyle_{o}} n{\scriptstyle_{o}} m{\scriptstyle_{e}}}$, where B${\scriptstyle_{o}}$ is the ambient magnetic field magnitude, $\mu{\scriptstyle_{o}}$ the permeability of free space, n${\scriptstyle_{o}}$ the background plasma number density, and m${\scriptstyle_{e}}$ the electron mass.  However, the beams rarely exceed 6000 km/s and the corresponding V${\scriptstyle_{Ae}}$ is typically $\gtrsim$ 50,000 km/s.  Thus, it is unlikely that the observed beam/flattop-related whistlers are driven unstable by the mechanism proposed by \citet{sauer10a}.  We also note that the electron distributions used in this study have energies almost entirely below the energies expected by \citet{sauer10a}.  Thus, a beam may exist above 50,000 km/s but we are unable to observe it in the EESA distributions.
\section{Results and Discussion}  \label{sec:theory}
\indent  We present a statistical study of the properties of large amplitude whistlers and comparison of wave amplitudes to AE-index to further characterize the waves described by \citet{kellogg10b}.  We have also presented the largest amplitude ($\gtrsim$8 nT peak-to-peak) whistler wave measured by a search coil in the radiation belts.  The highest occurrence probability was within $\pm$6 hours of midnight in the radiation belts between 3 $\leq$ L $\leq$ 6.  In addition, we describe the dependence of whistler characteristics on the shape of the electron distributions for energies below $\sim$30 keV.  However, one should note that because the Wind mission was not focused on the magnetosphere, the coverage is limited.  \\
\indent  We also show that very large amplitude whistlers are common in the radiation belts, consistent with recent results \citep{cattell08, cully08a, kellogg10b} using high time resolution field detectors.  Many previous studies that focused on time-averaged spectral intensity plots underestimated the amplitude of these bursty whistlers while the satellites with waveform capture capacities have only recently made measurements in the radiation belts.  In addition to their common occurrence, simulations have shown that these waves could be capable of producing significant changes in the relativistic electron fluxes of the radiation belts \citep{cattell08, bortnik08a}.  \\
\indent  The majority of the waveforms observed in our study have f/f${\scriptstyle_{ce}}$ $\leq$ 0.5, occurred within L $=$ 6, simultaneously with anisotropic electron distribution functions below 30 keV, and the majority of the largest amplitude whistlers observed occur during magnetically active periods (AE $>$ 200 nT), consistent with recent observations \citep{li10a}.  The correlation with AE suggests the source of free energy may be due to plasma injections from the geomagnetic tail.  Examination of LANL low energy (30-300 keV) electron summary data (not shown) showed large injections on the nightside at geosynchronous orbit shortly before Wind observed the two whistlers in Figure \ref{fig:whistlers}, supporting the idea that plasma injections may provide some of the free energy for large amplitude whistler generation.  Because the SST detectors were designed for the solar wind, we could not identify features in the high energy ($>$70 keV) electrons associated with the observed whistlers.  Regardless of what the source of free energy may be for the observed waves, we observe differences in wave properties depending on the type of electron distribution functions they are observed with.  \\
\indent  The whistlers associated with anisotropic electron distribution functions in the radiation belts exhibited broader frequency peaks than the whistlers associated with beam/flattop distributions in the outer magnetosphere.  For the 45 TDSS samples in which we were able to determine $\theta{\scriptstyle_{kB}}$, we found that the waves are often very oblique with $\theta{\scriptstyle_{kB}}$'s ranging from 0$^{\circ}$ $\leq$ $\theta{\scriptstyle_{kB}}$ $<$ 90$^{\circ}$, occasionally more oblique than the whistlers reported by \citet{cattell08} and \citet{kellogg10b}.  The anisotropy-related whistlers in both regions were more oblique than the beam/flattop-related whistlers observed in the outer magnetosphere.  The differences in propagation angle and width of frequency peak may be due to differences in the sources of free energy.  Some theories suggest that whistlers are generated with parallel wave vectors near the magnetic equator and they become more oblique as they propagate to higher latitudes.  So we examined the relationship between $\theta{\scriptstyle_{kB}}$ and $\lambda{\scriptstyle_{GSM}}$.  We found no correlation between $\theta{\scriptstyle_{kB}}$ and $\lambda{\scriptstyle_{GSM}}$, so the waves do not appear to be produced as parallel waves which propagate to the satellite and become more oblique as they propagate due to dispersive effects.  The anisotropy-related whistlers in both the radiation belts and outer magnetosphere are highly oblique, which results in a large longitudinal electric field.  When this electric field becomes sufficiently large, the typical cyclotron resonances can become less important than a Cerenkov resonance \citep{kennel66, kellogg10b}.  Recent studies have found that anisotropic bi-Maxwellian electron distributions can produce whistlers with peak growth rates for $\theta{\scriptstyle_{kB}}$ $\sim$ 50$^{\circ}$, suggesting that oblique whistlers may be driven by the anisotropic electron distributions observed \citep{hashimoto81a, schriver10a, kellogg10b}.  \\
\indent  Though we observe distinct differences in the whistler characteristics for different electron distribution functions, we cannot definitively state that the observed features are the source of free energy for these waves.  \citet{kellogg10b} presented a warm plasma instability analysis for one anisotropic electron distribution function measured by EESA Low near the whistler observed in Figure 4 on the right finding the distribution to be unstable.  The analysis showed that maximum growth occurred at real frequencies corresponding to the observed wave frequency, but the growth rate (0.08 s) and propagation angle ($\sim$10$^{\circ}$) were too low.  The instability was found to be a Cerenkov resonance \citep{kennel66} near 150 eV, not the typical cyclotron resonance, due to the highly oblique nature of the waves.  Owing to the long duration necessary for the EESA Low instrument to collect a full 4$\pi$ steradian distribution, one can assume that the instantaneous distributions are less isotropic.  Thus, \citet{kellogg10b} performed a second instability analysis with a more anisotropic hot electron component finding the growth rate (1.5 ms) and propagation angle (up to $\sim$60$^{\circ}$) to be approximately the same as the measured values (they measured $\theta{\scriptstyle_{kB}}$ $\sim$ 61$^{\circ}$).  Although, the real frequency produced by the analysis did not match the observed frequency.  They definitively showed that the distribution was unstable to whistler waves through a Cerenkov resonance.  Meaning, the parallel Landau resonant energies estimated by Equation \ref{eq:paracycenergy} fall well within the energy range of the EESA detectors.  
\section{Conclusions}  \label{sec:conclusions}
\indent  To summarize, we observed 244(65) whistlers inside L $=$ 15 using electric(magnetic) field data from the Wind spacecraft; $\sim$40$\%$($\sim$62$\%$) of the waves have peak-to-peak amplitudes of $\geq$50 mV/m($\geq$0.5 nT).  The majority of the whistlers had f/f${\scriptstyle_{ce}}$ $\leq$ 0.5 and were observed within $\pm$6 hours of midnight between 3 $\leq$ L $\leq$ 6.  The waves had a broad range of propagation angles, 0$^{\circ}$ $\leq$ $\theta{\scriptstyle_{kB}}$ $<$ 90$^{\circ}$, and $\theta{\scriptstyle_{kB}}$ was not correlated with $\lambda{\scriptstyle_{GSM}}$.  Thus, assuming an equatorial source, the wave propagation angles are not a consequence of propagation effects.  \\
\indent  One whistler observed by the Wind search coil had a wave amplitude over two orders of magnitude above previous observations, large enough to saturate the detector.  Also, we only measured one component of the magnetic field magnitude for this wave.  Thus, the wave amplitude of $\sim$8 nT peak-to-peak and our Poynting flux estimate of $\sim$300 $\mu$W/m$^{2}$ are underestimates of the true values.  A quasi-linear calculation of the time scales necessary for a wave with this amplitude to accelerate plasma sheet electrons to $\sim$1 MeV, assuming 1$\%$ efficiency, yields a time scale of roughly 0.5 seconds, much shorter than previous estimates based on small amplitude whistlers which were on the order of days.  Using the estimates of maximum change in kinetic energy from \citet{omura07a}, we found the 8 nT whistler to be capable of producing $\sim$61 MeV electrons through relativistic turning acceleration.  \\
\indent  Although we were not able to determine a source of free energy for the observed waves there were physical differences in the whistlers associated with different electron distributions.  In addition to the anisotropy-related whistlers having larger $\Delta$f/f than the rest of the whistlers, we observed differences in wave characteristics with the following electron parameters:  T${\scriptstyle_{\perp, EL}}$/T${\scriptstyle_{\parallel, EL}}$, T${\scriptstyle_{\perp, EH}}$/T${\scriptstyle_{\parallel, EH}}$, T${\scriptstyle_{EL}}$, and T${\scriptstyle_{EH}}$ (where EL $=$ EESA Low and EH $=$ EESA High).  The whistlers observed in the radiation belts simultaneously with anisotropic electron distributions had larger $\Delta$f/f than the rest of the whistlers observed.  We also examined the LANL low energy (30-300 keV) electron summary data and found that our largest events were associated with large substorm injections, consistent with the increase in wave amplitudes with increasing AE-index.  Our study adds to the mounting evidence that very large amplitude whistler waves are an important phenomena in radiation belt dynamics.  

\section{acknowledgments}  \label{sec:acknowledgments}
  We thank R. Lin (3DP), K. Ogilvie (SWE), R. Lepping (MFI), and E. Dors (LANL data) for the use of data from their instruments.  We would also like to thank M. Pulupa, S.D. Bale, and P. Schroeder for technical help with the 3DP software and analysis.  We thank L. Wang for help in calibration of the SST Foil data.  The authors thank World Data Center for Geomagnetism, Kyoto, for providing the AE index.  This research was supported by NESSF grant NNX07AU72H, grant NNX07AI05G, the Dr. Leonard Burlaga/Arctowski Medal Fellowship, and a contract from APL for the development of RBSP/EFW. \\


\begin{thebibliography}{39}
\providecommand{\natexlab}[1]{#1}
\expandafter\ifx\csname urlstyle\endcsname\relax
  \providecommand{\doi}[1]{doi:\discretionary{}{}{}#1}\else
  \providecommand{\doi}{doi:\discretionary{}{}{}\begingroup
  \urlstyle{rm}\Url}\fi

\bibitem[{\textit{{Bertucci} et~al.}(2005)\textit{{Bertucci}, {Achilleos},
  {Russell}, {Dougherty}, {Smith}, {Burton}, {Tsurutani}, and
  {Mazelle}}}]{bertucci05a}
{Bertucci}, C., N.~{Achilleos}, C.~T. {Russell}, M.~K. {Dougherty}, E.~J.
  {Smith}, M.~{Burton}, B.~T. {Tsurutani}, and C.~{Mazelle} (2005), {Bow Shock
  and Upstream Waves at Jupiter and Saturn: Cassini Magnetometer Observations},
  in \textit{The Physics of Collisionless Shocks: 4th Annual IGPP International
  Astrophysics Conference}, \textit{American Institute of Physics Conference
  Series}, vol. 781, edited by G.~{Li}, G.~P. {Zank}, and C.~T. {Russell}, pp.
  109--115, \doi{10.1063/1.2032682}.

\bibitem[{\textit{{Bortnik} et~al.}(2006)\textit{{Bortnik}, {Inan}, and
  {Bell}}}]{bortnik06a}
{Bortnik}, J., U.~S. {Inan}, and T.~F. {Bell} (2006), {Landau damping and
  resultant unidirectional propagation of chorus waves}, \textit{Geophys. Res.
  Lett.}, \textit{33}, 3102--+, \doi{10.1029/2005GL024553}.

\bibitem[{\textit{{Bortnik} et~al.}(2008)\textit{{Bortnik}, {Thorne}, and
  {Inan}}}]{bortnik08a}
{Bortnik}, J., R.~M. {Thorne}, and U.~S. {Inan} (2008), {Nonlinear interaction
  of energetic electrons with large amplitude chorus}, \textit{Geophys. Res.
  Lett.}, \textit{35}, 21,102--+, \doi{10.1029/2008GL035500}.

\bibitem[{\textit{{Bougeret} et~al.}(1995)\textit{{Bougeret}, {Kaiser},
  {Kellogg}, {Manning}, {Goetz}, {Monson}, {Monge}, {Friel}, {Meetre},
  {Perche}, {Sitruk}, and {Hoang}}}]{bougeret95}
{Bougeret}, J.-L., M.~L. {Kaiser}, P.~J. {Kellogg}, R.~{Manning}, K.~{Goetz},
  S.~J. {Monson}, N.~{Monge}, L.~{Friel}, C.~A. {Meetre}, C.~{Perche},
  L.~{Sitruk}, and S.~{Hoang} (1995), {Waves: The Radio and Plasma Wave
  Investigation on the Wind Spacecraft}, \textit{Space Science Reviews},
  \textit{71}, 231--263, \doi{10.1007/BF00751331}.

\bibitem[{\textit{{Breneman} et~al.}(2010)\textit{{Breneman}, {Cattell},
  {Schreiner}, {Kersten}, {Wilson III}, {Kellogg}, {Goetz}, and
  {Jian}}}]{breneman10a}
{Breneman}, A.~W., C.~A. {Cattell}, S.~{Schreiner}, K.~{Kersten}, L.~B. {Wilson
  III}, P.~J. {Kellogg}, K.~{Goetz}, and L.~K. {Jian} (2010), {Observations of
  Large Amplitude, Narrowband Whistlers at Stream Interaction Regions},
  \textit{J. Geophys. Res.}, \textit{115}, 8104--+, \doi{10.1029/2009JA014920}.

\bibitem[{\textit{{Burtis} and {Helliwell}}(1969)}]{burtis69a}
{Burtis}, W.~J., and R.~A. {Helliwell} (1969), {Banded chorus - A new type of
  VLF radiation observed in the magnetosphere by OGO 1 and OGO 3.}, \textit{J.
  Geophys. Res.}, \textit{74}, 3002--3010, \doi{10.1029/JA074i011p03002}.

\bibitem[{\textit{{Cattell} et~al.}(2008)\textit{{Cattell}, {Wygant}, {Goetz},
  {Kersten}, {Kellogg}, {von Rosenvinge}, {Bale}, {Roth}, {Temerin}, {Hudson},
  {Mewaldt}, {Wiedenbeck}, {Maksimovic}, {Ergun}, {Acuna}, and
  {Russell}}}]{cattell08}
{Cattell}, C., J.~R. {Wygant}, K.~{Goetz}, K.~{Kersten}, P.~J. {Kellogg},
  T.~{von Rosenvinge}, S.~D. {Bale}, I.~{Roth}, M.~{Temerin}, M.~K. {Hudson},
  R.~A. {Mewaldt}, M.~{Wiedenbeck}, M.~{Maksimovic}, R.~{Ergun}, M.~{Acuna},
  and C.~T. {Russell} (2008), {Discovery of very large amplitude whistler-mode
  waves in Earth's radiation belts}, \textit{Geophys. Res. Lett.}, \textit{35},
  1105--+, \doi{10.1029/2007GL032009}.

\bibitem[{\textit{{Cully} et~al.}(2008)\textit{{Cully}, {Bonnell}, and
  {Ergun}}}]{cully08a}
{Cully}, C.~M., J.~W. {Bonnell}, and R.~E. {Ergun} (2008), {THEMIS observations
  of long-lived regions of large-amplitude whistler waves in the inner
  magnetosphere}, \textit{Geophys. Res. Lett.}, \textit{35}, 17--+,
  \doi{10.1029/2008GL033643}.

\bibitem[{\textit{{Gary} et~al.}(1994)\textit{{Gary}, {Scime}, {Phillips}, and
  {Feldman}}}]{gary94}
{Gary}, S.~P., E.~E. {Scime}, J.~L. {Phillips}, and W.~C. {Feldman} (1994),
  {The whistler heat flux instability: Threshold conditions in the solar wind},
  \textit{J. Geophys. Res.}, \textit{99}, 23,391--+, \doi{10.1029/94JA02067}.

\bibitem[{\textit{{Gurnett} et~al.}(1976)\textit{{Gurnett}, {Frank}, and
  {Lepping}}}]{gurnett76a}
{Gurnett}, D.~A., L.~A. {Frank}, and R.~P. {Lepping} (1976), {Plasma waves in
  the distant magnetotail}, \textit{J. Geophys. Res.}, \textit{81}, 6059--6071,
  \doi{10.1029/JA081i034p06059}.

\bibitem[{\textit{{Hashimoto} and {Kimura}}(1981)}]{hashimoto81a}
{Hashimoto}, K., and I.~{Kimura} (1981), {A generation mechanism of narrow band
  hiss emissions above one half the electron cyclotron frequency in the outer
  magnetosphere}, \textit{J. Geophys. Res.}, \textit{86}, 11,148--11,152,
  \doi{10.1029/JA086iA13p11148}.

\bibitem[{\textit{{Hoppe} et~al.}(1981)\textit{{Hoppe}, {Russell}, {Frank},
  {Eastman}, and {Greenstadt}}}]{hoppe81a}
{Hoppe}, M.~M., C.~T. {Russell}, L.~A. {Frank}, T.~E. {Eastman}, and E.~W.
  {Greenstadt} (1981), {Upstream hydromagnetic waves and their association with
  backstreaming ion populations - ISEE 1 and 2 observations}, \textit{J.
  Geophys. Res.}, \textit{86}, 4471--4492, \doi{10.1029/JA086iA06p04471}.

\bibitem[{\textit{{Horne} et~al.}(2003)\textit{{Horne}, {Glauert}, and
  {Thorne}}}]{horne03a}
{Horne}, R.~B., S.~A. {Glauert}, and R.~M. {Thorne} (2003), {Resonant diffusion
  of radiation belt electrons by whistler-mode chorus}, \textit{Geophys. Res.
  Lett.}, \textit{30}, 090,000--1, \doi{10.1029/2003GL016963}.

\bibitem[{\textit{{Horne} et~al.}(2005)\textit{{Horne}, {Thorne}, {Glauert},
  {Albert}, {Meredith}, and {Anderson}}}]{horne05a}
{Horne}, R.~B., R.~M. {Thorne}, S.~A. {Glauert}, J.~M. {Albert}, N.~P.
  {Meredith}, and R.~R. {Anderson} (2005), {Timescale for radiation belt
  electron acceleration by whistler mode chorus waves}, \textit{J. Geophys.
  Res.}, \textit{110}, 3225--+, \doi{10.1029/2004JA010811}.

\bibitem[{\textit{{Kellogg} et~al.}(2010)\textit{{Kellogg}, {Cattell}, {Goetz},
  {Monson}, and {Wilson}~III}}]{kellogg10b}
{Kellogg}, P.~J., C.~A. {Cattell}, K.~{Goetz}, S.~J. {Monson}, and L.~B.
  {Wilson}~III (2010), {Large Amplitude Whistlers in the Magnetosphere Observed
  with Wind-Waves}, \textit{J. Geophys. Res.}, submitted.

\bibitem[{\textit{{Kennel} and {Engelmann}}(1966)}]{kennel66b}
{Kennel}, C.~F., and F.~{Engelmann} (1966), {Velocity Space Diffusion from Weak
  Plasma Turbulence in a Magnetic Field}, \textit{Phys. Fluids}, \textit{9},
  2377--2388, \doi{10.1063/1.1761629}.

\bibitem[{\textit{{Kennel} and {Petscheck}}(1966)}]{kennel66}
{Kennel}, C.~F., and H.~E. {Petscheck} (1966), {Limit on stably trapped
  particle fluxes}, \textit{J. Geophys. Res.}, \textit{71}, 1--28.

\bibitem[{\textit{{Khrabrov} and {Sonnerup}}(1998)}]{khrabrov98}
{Khrabrov}, A.~V., and B.~U.~{\"O}. {Sonnerup} (1998), {Error estimates for
  minimum variance analysis}, \textit{J. Geophys. Res.}, \textit{103},
  6641--6652, \doi{10.1029/97JA03731}.

\bibitem[{\textit{{Kumagai} et~al.}(1980)\textit{{Kumagai}, {Hashimoto},
  {Kimura}, and {Matsumoto}}}]{kumagai80a}
{Kumagai}, H., K.~{Hashimoto}, I.~{Kimura}, and H.~{Matsumoto} (1980),
  {Computer simulation of a Cerenkov interaction between obliquely propagating
  whistler mode waves and an electron beam}, \textit{Phys. Fluids},
  \textit{23}, 184--193, \doi{10.1063/1.862837}.

\bibitem[{\textit{{Le Contel} et~al.}(2009)\textit{{Le Contel}, {Roux},
  {Jacquey}, {Robert}, {Berthomier}, {Chust}, {Grison}, {Angelopoulos},
  {Sibeck}, {Chaston}, {Cully}, {Ergun}, {Glassmeier}, {Auster}, {McFadden},
  {Carlson}, {Larson}, {Bonnell}, {Mende}, {Russell}, {Donovan}, {Mann}, and
  {Singer}}}]{lecontel09a}
{Le Contel}, O., A.~{Roux}, C.~{Jacquey}, P.~{Robert}, M.~{Berthomier},
  T.~{Chust}, B.~{Grison}, V.~{Angelopoulos}, D.~{Sibeck}, C.~C. {Chaston},
  C.~M. {Cully}, B.~{Ergun}, K.~{Glassmeier}, U.~{Auster}, J.~{McFadden},
  C.~{Carlson}, D.~{Larson}, J.~W. {Bonnell}, S.~{Mende}, C.~T. {Russell},
  E.~{Donovan}, I.~{Mann}, and H.~{Singer} (2009), {Quasi-parallel whistler
  mode waves observed by THEMIS during near-earth dipolarizations},
  \textit{Ann. Geophys.}, \textit{27}, 2259--2275.

\bibitem[{\textit{{Lepping} et~al.}(1995)\textit{{Lepping}, {Ac{\~u}na},
  {Burlaga}, {Farrell}, {Slavin}, {Schatten}, {Mariani}, {Ness}, {Neubauer},
  {Whang}, {Byrnes}, {Kennon}, {Panetta}, {Scheifele}, and
  {Worley}}}]{lepping95}
{Lepping}, R.~P., M.~H. {Ac{\~u}na}, L.~F. {Burlaga}, W.~M. {Farrell}, J.~A.
  {Slavin}, K.~H. {Schatten}, F.~{Mariani}, N.~F. {Ness}, F.~M. {Neubauer},
  Y.~C. {Whang}, J.~B. {Byrnes}, R.~S. {Kennon}, P.~V. {Panetta},
  J.~{Scheifele}, and E.~M. {Worley} (1995), {The Wind Magnetic Field
  Investigation}, \textit{Space Science Reviews}, \textit{71}, 207--229,
  \doi{10.1007/BF00751330}.

\bibitem[{\textit{{Li} et~al.}(2010)\textit{{Li}, {Thorne}, {Nishimura},
  {Bortnik}, {Angelopoulos}, {McFadden}, {Larson}, {Bonnell}, {LeContel},
  {Roux}, and {Auster}}}]{li10a}
{Li}, W., R.~M. {Thorne}, Y.~{Nishimura}, J.~{Bortnik}, V.~{Angelopoulos},
  J.~P. {McFadden}, D.~E. {Larson}, J.~W. {Bonnell}, O.~{LeContel}, A.~{Roux},
  and U.~{Auster} (2010), {THEMIS analysis of observed equatorial electron
  distributions responsible for the chorus excitation}, \textit{J. Geophys.
  Res.}, \textit{115}, A00F11, \doi{10.1029/2009JA014845}.

\bibitem[{\textit{{Lin} et~al.}(1995)\textit{{Lin}, {Anderson}, {Ashford},
  {Carlson}, {Curtis}, {Ergun}, {Larson}, {McFadden}, {McCarthy}, {Parks},
  {R{\`e}me}, {Bosqued}, {Coutelier}, {Cotin}, {D'Uston}, {Wenzel},
  {Sanderson}, {Henrion}, {Ronnet}, and {Paschmann}}}]{lin95}
{Lin}, R.~P., K.~A. {Anderson}, S.~{Ashford}, C.~{Carlson}, D.~{Curtis},
  R.~{Ergun}, D.~{Larson}, J.~{McFadden}, M.~{McCarthy}, G.~K. {Parks},
  H.~{R{\`e}me}, J.~M. {Bosqued}, J.~{Coutelier}, F.~{Cotin}, C.~{D'Uston},
  K.-P. {Wenzel}, T.~R. {Sanderson}, J.~{Henrion}, J.~C. {Ronnet}, and
  G.~{Paschmann} (1995), {A Three-Dimensional Plasma and Energetic Particle
  Investigation for the Wind Spacecraft}, \textit{Space Science Reviews},
  \textit{71}, 125--153, \doi{10.1007/BF00751328}.

\bibitem[{\textit{{Lyons} et~al.}(1972)\textit{{Lyons}, {Thorne}, and
  {Kennel}}}]{lyons72a}
{Lyons}, L.~R., R.~M. {Thorne}, and C.~F. {Kennel} (1972), {Pitch-angle
  diffusion of radiation belt electrons within the plasmasphere.}, \textit{J.
  Geophys. Res.}, \textit{77}, 3455--3474, \doi{10.1029/JA077i019p03455}.

\bibitem[{\textit{{Meredith} et~al.}(2001)\textit{{Meredith}, {Horne}, and
  {Anderson}}}]{meredith01a}
{Meredith}, N.~P., R.~B. {Horne}, and R.~R. {Anderson} (2001), {Substorm
  dependence of chorus amplitudes: Implications for the acceleration of
  electrons to relativistic energies}, \textit{J. Geophys. Res.}, \textit{106},
  13,165--13,178, \doi{10.1029/2000JA900156}.

\bibitem[{\textit{{Neubauer} and {Musmann}}(1977)}]{neubauer77a}
{Neubauer}, F.~M., and G.~{Musmann} (1977), {Fast magnetic fluctuations in the
  solar wind - HELIOS I}, \textit{J. Geophys. Res.}, \textit{82}, 3201--3212,
  \doi{10.1029/JA082i022p03201}.

\bibitem[{\textit{{Omura} et~al.}(2007)\textit{{Omura}, {Furuya}, and
  {Summers}}}]{omura07a}
{Omura}, Y., N.~{Furuya}, and D.~{Summers} (2007), {Relativistic turning
  acceleration of resonant electrons by coherent whistler mode waves in a
  dipole magnetic field}, \textit{J. Geophys. Res.}, \textit{112}, 6236--+,
  \doi{10.1029/2006JA012243}.

\bibitem[{\textit{{Parrot} et~al.}(2003)\textit{{Parrot}, {Santol{\'y}k},
  {Cornilleau-Wehrlin}, {Maksimovic}, and {Harvey}}}]{parrot03a}
{Parrot}, M., O.~{Santol{\'y}k}, N.~{Cornilleau-Wehrlin}, M.~{Maksimovic}, and
  C.~C. {Harvey} (2003), {Source location of chorus emissions observed by
  Cluster}, \textit{Ann. Geophys.}, \textit{21}, 473--480.

\bibitem[{\textit{{Russell} et~al.}(1969)\textit{{Russell}, {Holzer}, and
  {Smith}}}]{russell69a}
{Russell}, C.~T., R.~E. {Holzer}, and E.~J. {Smith} (1969), {OGO 3 observations
  of ELF noise in the magnetosphere. 1. Spatial extent and frequency of
  occurrence.}, \textit{J. Geophys. Res.}, \textit{74}, 755--777,
  \doi{10.1029/JA074i003p00755}.

\bibitem[{\textit{{Santol{\'{\i}}k} et~al.}(2003)\textit{{Santol{\'{\i}}k},
  {Gurnett}, {Pickett}, {Parrot}, and {Cornilleau-Wehrlin}}}]{santolik03a}
{Santol{\'{\i}}k}, O., D.~A. {Gurnett}, J.~S. {Pickett}, M.~{Parrot}, and
  N.~{Cornilleau-Wehrlin} (2003), {Spatio-temporal structure of storm-time
  chorus}, \textit{J. Geophys. Res.}, \textit{108}, 1278--+,
  \doi{10.1029/2002JA009791}.

\bibitem[{\textit{{Santol{\'{\i}}k} et~al.}(2010)\textit{{Santol{\'{\i}}k},
  {Pickett}, {Gurnett}, {Menietti}, {Tsurutani}, and
  {Verkhoglyadova}}}]{santolik10a}
{Santol{\'{\i}}k}, O., J.~S. {Pickett}, D.~A. {Gurnett}, J.~D. {Menietti},
  B.~T. {Tsurutani}, and O.~{Verkhoglyadova} (2010), {Survey of Poynting flux
  of whistler mode chorus in the outer zone}, \textit{J. Geophys. Res.},
  \textit{115}, 0--+, \doi{10.1029/2009JA014925}.

\bibitem[{\textit{{Sauer} and {Sydora}}(2010)}]{sauer10a}
{Sauer}, K., and R.~D. {Sydora} (2010), {Beam-excited whistler waves at oblique
  propagation with relation to STEREO radiation belt observations},
  \textit{Ann. Geophys.}, \textit{28}, 1317--1325.

\bibitem[{\textit{{Schriver} et~al.}(2010)\textit{{Schriver}, {Ashour-Abdalla},
  {Coroniti}, {LeBoeuf}, {Decyk}, {Travnicek}, {Santol{\'{\i}}k}, {Winningham},
  {Pickett}, {Goldstein}, and {Fazakerley}}}]{schriver10a}
{Schriver}, D., M.~{Ashour-Abdalla}, F.~V. {Coroniti}, J.~N. {LeBoeuf},
  V.~{Decyk}, P.~{Travnicek}, O.~{Santol{\'{\i}}k}, D.~{Winningham}, J.~S.
  {Pickett}, M.~L. {Goldstein}, and A.~N. {Fazakerley} (2010), {Generation of
  whistler mode emissions in the inner magnetosphere: An event study},
  \textit{J. Geophys. Res.}, \textit{115}, A00F17--+,
  \doi{10.1029/2009JA014932}.

\bibitem[{\textit{{Singh}}(1972)}]{singh72a}
{Singh}, R.~P. (1972), {Amplification of signal by Cerenkov resonance
  interaction}, \textit{Planet. Space Sci.}, \textit{20}, 2073--+,
  \doi{10.1016/0032-0633(72)90063-3}.

\bibitem[{\textit{{Starodubtsev} et~al.}(1999)\textit{{Starodubtsev}, {Krafft},
  {Lundin}, and {Th{\'e}venet}}}]{starodubtsev99c}
{Starodubtsev}, M., C.~{Krafft}, B.~{Lundin}, and P.~{Th{\'e}venet} (1999),
  {Resonant Cherenkov emission of whistlers by a modulated electron beam},
  \textit{Phys. Plasmas}, \textit{6}, 2862--2869, \doi{10.1063/1.873244}.

\bibitem[{\textit{{Tokar} et~al.}(1984)\textit{{Tokar}, {Gurnett}, and
  {Feldman}}}]{tokar84}
{Tokar}, R.~L., D.~A. {Gurnett}, and W.~C. {Feldman} (1984), {Whistler mode
  turbulence generated by electron beams in earth's bow shock}, \textit{J.
  Geophys. Res.}, \textit{89}, 105--114, \doi{10.1029/JA089iA01p00105}.

\bibitem[{\textit{{Tsurutani} et~al.}(1987)\textit{{Tsurutani}, {Smith},
  {Thorne}, {Gosling}, and {Matsumoto}}}]{tsurutani87}
{Tsurutani}, B.~T., E.~J. {Smith}, R.~M. {Thorne}, J.~T. {Gosling}, and
  H.~{Matsumoto} (1987), {Steepened magnetosonic waves at Comet
  Giacobini-Zinner}, \textit{J. Geophys. Res.}, \textit{92}, 11,074--11,082,
  \doi{10.1029/JA092iA10p11074}.

\bibitem[{\textit{{Wilson III} et~al.}(2009)\textit{{Wilson III}, {Cattell},
  {Kellogg}, {Goetz}, {Kersten}, {Kasper}, {Szabo}, and
  {Meziane}}}]{wilsoniii09a}
{Wilson III}, L.~B., C.~A. {Cattell}, P.~J. {Kellogg}, K.~{Goetz},
  K.~{Kersten}, J.~C. {Kasper}, A.~{Szabo}, and K.~{Meziane} (2009),
  {Low-frequency whistler waves and shocklets observed at quasi-perpendicular
  interplanetary shocks}, \textit{J. Geophys. Res.}, \textit{114}, 10,106--+,
  \doi{10.1029/2009JA014376}.

\bibitem[{\textit{{Zhang} et~al.}(1993)\textit{{Zhang}, {Matsumoto}, and
  {Omura}}}]{zhang93a}
{Zhang}, Y.~L., H.~{Matsumoto}, and Y.~{Omura} (1993), {Linear and nonlinear
  interactions of an electron beam with oblique whistler and electrostatic
  waves in the magnetosphere}, \textit{J. Geophys. Res.}, \textit{98},
  21,353--+, \doi{10.1029/93JA01937}.

\end{thebibliography}

\newpage

\begin{table}[htb]
  \caption{Wind Petal Orbit Whistler Statistics}
  \label{tab:whistlerstats}
    \begin{tabular}{| c | c | c | c | c |}
      \hline  \hline
    \multicolumn{5}{|c|}{\textbf{Whistler Location and Particle Statistics (L $<$ 15)}} \\
      \hline
    \multicolumn{3}{|c|}{Electron Distribution Function} & Inside of       & Between Magnetopause   \\
    \multicolumn{3}{|c|}{Type} & Radiation Belts & and Radiation Belts    \\
      \hline
      \multicolumn{3}{|c|}{Anisotropic}   &       172       &        42         \\
      \multicolumn{3}{|c|}{Beam/Flattop}  &         2       &        28         \\
      \hline
    \multicolumn{5}{|c|}{\textbf{Whistler Amplitude Statistics (L $<$ 15)}}  \\
      \hline
      \multicolumn{2}{|c|}{Whistler Waves with $\mid$\textbf{E}${\scriptstyle_{w}} \mid$} & $\geq$100 mV/m & $\geq$50 mV/m & $\geq$30 mV/m   \\
      \multicolumn{2}{|c|}{$\#$/(Total $\#$ measured with $\mid$\textbf{E}${\scriptstyle_{w}} \mid$)} &   48/244    &   97/244   &  129/244              \\
      \multicolumn{2}{|c|}{Range of L-Shells} &   3.8-5.8   &   3.7-6.3  &  3.7-7.9              \\
      \hline
      \multicolumn{2}{|c|}{Whistler Waves with $\mid$\textbf{B}${\scriptstyle_{w}} \mid$} & $\geq$1.0 nT & $\geq$0.5 nT & $\geq$0.3 nT       \\
      \multicolumn{2}{|c|}{$\#$/(Total $\#$ measured with $\mid$\textbf{B}${\scriptstyle_{w}} \mid$)} &   17/65     &   40/65    &  44/65                \\
      \multicolumn{2}{|c|}{Range of L-Shells} &   4.0-8.8   &  4.0-11.3  & 4.0-11.3              \\
      \hline
    \multicolumn{5}{|c|}{\textbf{Whistler AE (nT) Index Statistics (L $<$ 15)}}  \\
      \hline
      Type &  200 $\leq$ AE $\leq$ 400  &  400 $\leq$ AE $\leq$ 600  &  600 $\leq$ AE $\leq$ 800  &  800 $\leq$ AE $\leq$ 1000  \\
      \hline
      $\mid$\textbf{E}${\scriptstyle_{w}} \mid$ (mV/m) & 110.88 $\pm$ 6.45  & 100.56 $\pm$ 5.69  & 96.43 $\pm$ 4.75  & 40.35 $\pm$ 1.86   \\
      $\mid$\textbf{B}${\scriptstyle_{w}} \mid$ (nT)   &  1.61 $\pm$ 0.27  &   1.09 $\pm$ 0.04  &   1.65 $\pm$ 0.03  &   0.92 $\pm$ 0.03  \\
      $\#$/(Total $\#$ of $\mid$\textbf{E}${\scriptstyle_{w}} \mid$) & 34/217 & 60/217 & 14/217 & 35/217  \\
      $\#$/(Total $\#$ of $\mid$\textbf{B}${\scriptstyle_{w}} \mid$) & 13/52  &  9/52  & 2/52 & 13/52     \\
      \hline  \hline
    \end{tabular}
\end{table}

\end{document}